\documentclass[conference]{IEEEtran}
\IEEEoverridecommandlockouts

\usepackage{cite}
\usepackage{amsmath,amssymb,amsfonts}
\usepackage{algorithmic}
\usepackage{graphicx}
\usepackage{textcomp}
\usepackage{xcolor}
\usepackage{todonotes}
\usepackage{pgfplots}\pgfplotsset{compat=1.18}
\usepackage{subcaption}
\usepackage{booktabs}
\def\BibTeX{{\rm B\kern-.05em{\sc i\kern-.025em b}\kern-.08em
    T\kern-.1667em\lower.7ex\hbox{E}\kern-.125emX}}
\begin{document}

\title{End-to-End Optical Semantic Communication over a Nonlinear WDM Fiber Link [Invited]}

\author{
Hussein Jammal, Andrea Bianco, Cristina Rottondi
\\
\vspace{-1.0em}
\IEEEauthorblockA{Department of Electronics and Telecommunications, Politecnico di Torino, Turin, Italy}\\
\vspace{-1.0em}
\IEEEauthorblockA{%
\textit{name.surname@polito.it}
}
}
\author{
\IEEEauthorblockN{Hussein Jammal, Andrea Bianco, Cristina Rottondi}
\IEEEauthorblockA{
Department of Electronics and Telecommunications, Politecnico di Torino, Turin, Italy\\
\textit{name.surname@polito.it}
}
}

\definecolor{clr150}{HTML}{1B9E8F}
\definecolor{clr200}{HTML}{3FA85F}
\definecolor{clr300}{HTML}{76B33C}
\definecolor{clr400}{HTML}{9A9C24}
\definecolor{clr500}{HTML}{D79E2B}
\definecolor{clr600}{HTML}{E57D31}
\definecolor{clr700}{HTML}{D45536}
\definecolor{clr800}{HTML}{A82F2F}

\pgfplotsset{
  pwraxis/.style={
    width=\linewidth, height=4.8cm,
    xlabel={Launch power per channel [dBm]},
    xmin=-9.8, xmax=3.8, xtick={-9,-6,-3,-1,0,1,3},
    grid=both,
    grid style={line width=0.2pt, draw=gray!18},
    major grid style={line width=0.3pt, draw=gray!35},
    tick label style={font=\footnotesize},
    label style={font=\footnotesize},
    line width=1.0pt, mark size=2.2pt,
    legend style={font=\scriptsize, draw=none, fill=white, fill opacity=0.82,
                  text opacity=1, inner sep=2pt, row sep=-1pt},
    legend cell align=left,
  },
  mSemL/.style={mark=*,         solid},
  mSemF/.style={mark=o,         densely dashed},
  mLDPC/.style={mark=triangle*, solid},
  mBare/.style={mark=diamond,   densely dashdotted},
  c150/.style={color=clr150},
  c200/.style={color=clr200},
  c300/.style={color=clr300},
  c400/.style={color=clr400},
  c500/.style={color=clr500},
  c600/.style={color=clr600},
  c700/.style={color=clr700},
  c800/.style={color=clr800},
}

\newcommand{\lengthkey}{%
  \begin{tikzpicture}[baseline]
    \node[anchor=east, font=\footnotesize] at (0,0) {Link length [km]:};
    \foreach \c/\lab [count=\i from 0] in {clr150/150,
      clr200/200,
      clr300/300,
      clr400/400,
      clr500/500,
      clr600/600,
      clr700/700,
      clr800/800}
      {%
        \draw[color=\c, line width=1.2pt] (\i*1.28cm+0.22cm,0) -- ++(0.42cm,0);
        \node[anchor=west, font=\footnotesize] at (\i*1.28cm+0.68cm,0) {\lab};
      }
  \end{tikzpicture}%
}

\definecolor{clrM16}{HTML}{0072B2}
\definecolor{clrM64}{HTML}{E69F00}
\definecolor{clrM256}{HTML}{009E73}
\pgfplotsset{
  cM16/.style={color=clrM16},
  cM64/.style={color=clrM64},
  cM256/.style={color=clrM256},
  reachaxis/.style={
    width=\linewidth, height=4.8cm,
    xlabel={Link length [km]},
    xmin=130, xmax=820, xtick={150,200,300,400,500,600,700,800},
    grid=both,
    grid style={line width=0.2pt, draw=gray!18},
    major grid style={line width=0.3pt, draw=gray!35},
    tick label style={font=\footnotesize},
    label style={font=\footnotesize},
    line width=1.0pt, mark size=2.2pt,
  },
}
\newcommand{\modkey}{%
  \begin{tikzpicture}[baseline]
    \node[anchor=east, font=\footnotesize] at (0,0) {Modulation order:};
    \foreach \c/\lab [count=\i from 0] in {clrM16/{16-QAM}, clrM64/{64-QAM}, clrM256/{256-QAM}}
      {\draw[color=\c, line width=1.2pt] (\i*2.05cm+0.22cm,0) -- ++(0.5cm,0);
       \node[anchor=west, font=\footnotesize] at (\i*2.05cm+0.74cm,0) {\lab};}
  \end{tikzpicture}%
}
\newcommand{\schemesample}[2]{%
  \tikz[baseline=-0.6ex]{%
    \draw[line width=1.0pt,#1] (0,0)--(0.5,0);
    \draw[line width=1.0pt,#1,mark=#2,mark size=2.2pt,solid] plot coordinates {(0.25,0)};}%
}
\newcommand{\schemekey}{{\footnotesize%
  Scheme:\;\;
  \schemesample{solid}{*}\,~Proposed system (learned constellation)\quad
  \schemesample{densely dashed}{o}\,~Proposed system (fixed constellation)\quad
  \schemesample{solid}{triangle*}\,~LDPC-coded JPEG\quad
  \schemesample{densely dashdotted}{diamond}\,~Uncoded JPEG%
}}

\maketitle

\begin{abstract}
Emerging optical-network applications increasingly use received data for inference and control rather than exact source reproduction, creating an opportunity to trade bit-level fidelity for greater transmission reach and efficiency. We propose an end-to-end optical semantic communication system for joint image classification and reconstruction over a nonlinear wavelength-division multiplexed (WDM) fiber channel. The system maps each image directly into a fixed-length sequence of channel symbols that preserves task-relevant information, without explicit source compression or channel coding. Experiments on the MNIST dataset cover launch powers from $-9$ to $+3$~dBm, fiber lengths up to $800$~km, and $16$-, $64$-, and $256$-Quadrature Amplitude Modulation (QAM) formats. At $0$~dBm, classification accuracy remains between $98.92\%$ and $99.31\%$ across all tested link lengths and modulation orders, while requiring fewer transmitted symbols than a Low-Density Parity-Check (LDPC)-coded JPEG baseline at every tested modulation order. These results show that semantic communication can simultaneously extend optical reach and reduce transmission resources by conveying only task-relevant information.

\end{abstract}


\section{Introduction} \label{sec:introduction}
Many networked applications, including digital twins, distributed sensing, and machine-to-machine communication, require the delivery of information that is sufficient to accomplish a specific task rather than an exact reproduction of the source data. This task-oriented communication paradigm has become increasingly relevant with the growing use of artificial intelligence (AI), although the underlying requirement is not limited to AI-driven applications~\cite{Lan2021WhatIs, Xie2021DeepLearning, Qin2021Semantic}. As such applications increasingly rely on optical backbone, data-center, and access networks to transport their data, a mismatch emerges: the applications may only require task-relevant information, whereas conventional optical communication systems are designed for transparent and reliable recovery of the transmitted bitstream. 

Semantic communication addresses this mismatch through a compact \emph{semantic representation} that encodes the source data while preserving the information relevant to the end-to-end application task~\cite{Zhang2022Deep, Gunduz2022Beyond}. In joint source--channel coding (JSCC), the semantic encoder, channel-symbol mapping, and receiver are optimized together to maximize task performance after transmission~\cite{Bourtsoulatze2019Deep, Kurka2020Joint}. The resulting representation is therefore both task-aware and channel-aware. Changing the channel distribution used during optimization can change the semantic encoder parameters, the representation, and its mapping to transmitted symbols. This differs from a conventional pipeline in which source compression, forward-error correction (FEC), and modulation are designed independently for reliable bit recovery~\cite{Shannon1948Mathematical}.

In end-to-end JSCC, the joint optimization is typically performed by backpropagation and therefore requires a differentiable channel model to propagate gradients from the receiver to the transmitter. Since the channel model directly influences the learned semantic representation, it should also capture the relevant physical impairments as accurately as possible. This requirement is particularly challenging for coherent wavelength-division multiplexed (WDM) fiber links, where chromatic dispersion introduces memory, amplified spontaneous emission (ASE) noise accumulates across amplified spans, and Kerr nonlinearity makes the distortion dependent on the transmitted waveform and launch power~\cite{Agrawal2019Nonlinear}. Differentiable Gaussian approximations enable gradient-based training, but represent nonlinear propagation through aggregate noise statistics and may therefore miss temporal correlations and nonlinear symbol interactions~\cite{Poggiolini2026Polynomial}. Split-step Fourier method (SSFM) models capture these effects more directly, but do not readily provide the channel derivatives required for backpropagation. Consequently, end-to-end optical learning over SSFM models has mainly focused on modulation, bit or symbol recovery, and achievable information rate~\cite{Jovanovic2023Geometric,Jovanovic2022End,Gaiarin2020End}, whereas optical semantic JSCC has relied primarily on differentiable Gaussian approximations~\cite{Cai2026Convolutional,Cai2025Machine}. 
Thus, directly optimizing an optical semantic communication system over a nonlinear WDM fiber simulator without channel derivatives, and evaluating its robustness across launch powers and link lengths, remains an open problem~\cite{AitAoudia2018End}.

To address this gap, we propose and evaluate an optical semantic communication system optimized using a black-box, non-differentiable Manakov-SSFM model of a coherent nonlinear WDM fiber link. The proposed system performs joint image classification and reconstruction and maps the semantic representation onto channel symbols without any explicit source compression or channel coding stage. We consider two variants: one in which the constellation geometry is learned jointly with the transmitter and receiver, and one in which a conventional quadrature amplitude modulation (QAM) constellation is kept fixed. In both cases, the receiver processes soft channel outputs through a dual-head architecture for classification and reconstruction. To train the system, we adopt a two-phase strategy that first pretrains the components on a differentiable additive white Gaussian noise (AWGN) surrogate and then alternates gradient-based receiver updates with policy-gradient transmitter updates driven by task performance measured after fiber~\cite{Zhang2022Deep}; in the learned-constellation variant, the constellation is also updated during receiver optimization. We evaluate both variants across launch powers from $-9$ to $+3$~dBm, link lengths from $150$ to $800$~km, and QAM orders of $16$, $64$, and $256$, and compare them at equal launch power with conventional JPEG pipelines without FEC and with low-density parity-check (LDPC) coding. The proposed system preserves task performance across the tested conditions, degrades gradually where the conventional pipelines collapse, and transmits fewer symbols than the LDPC-coded JPEG pipeline.

The remainder of this paper is organized as follows: Sec.~\ref{sec:semantic_jscc} presents the system model and end-to-end optimization, Sec.~\ref{sec:experimental-protocol} describes the experimental setup, Sec.~\ref{sec:results} discusses the results and Sec.~\ref{sec:conclusion} concludes the paper.  

\begin{figure*}[t]
    \centering
    \includegraphics[
        width=0.89\textwidth,
        keepaspectratio
    ]{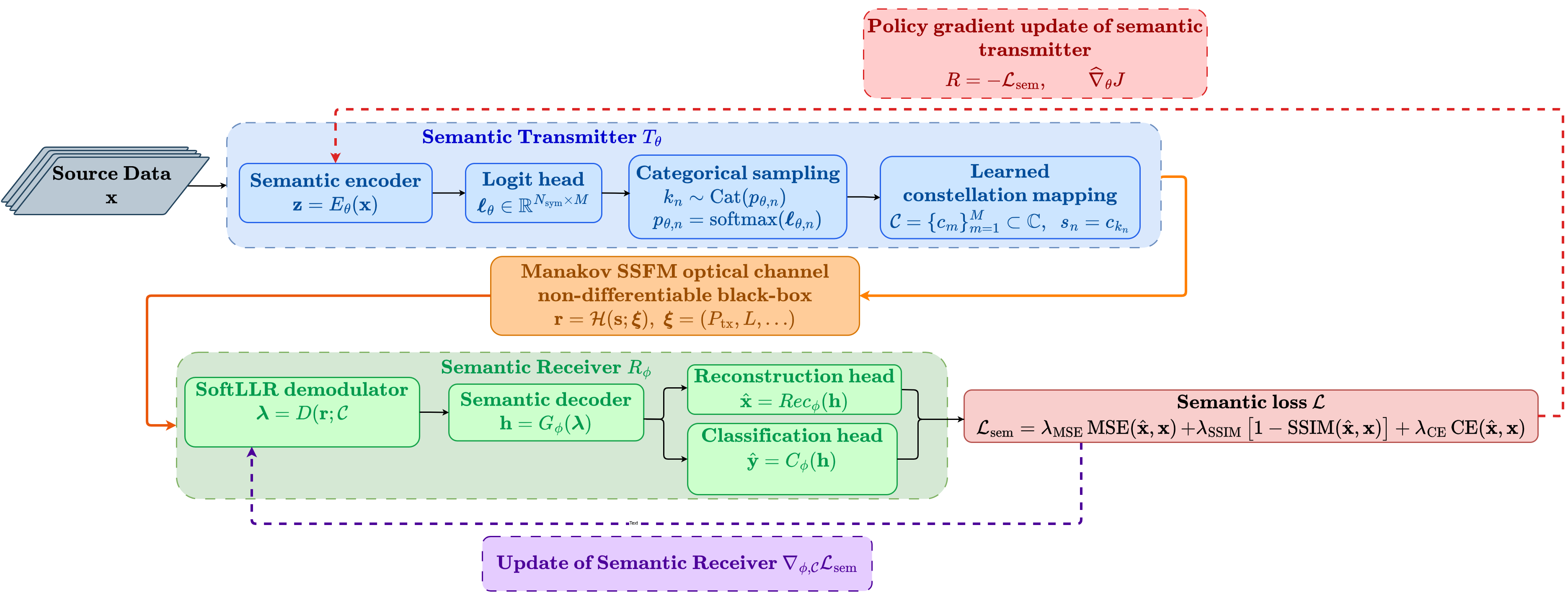}
    \caption{Architecture and optimization flow of the proposed system.}
    \label{fig:semcom_pipeline}
    \vspace{-1.5em}
\end{figure*}

\section{System Model and End-to-End Optimization}\label{sec:semantic_jscc}
\subsection{Problem Formulation}\label{subsec:problem-formulation}

Let $\mathcal{D}=\{(\mathbf{x}_i,y_i)\}_{i=1}^{N}$ denote a labeled image dataset, where $\mathbf{x}_i\in[0,1]^{H\times W}$ is an image and $y_i\in\{1,\ldots,K\}$ is its class label. The semantic transmitter $T_{\theta}$, with trainable parameters $\theta$, maps $\mathbf{x}$ to a block $\mathbf{s}\in\mathbb{C}^{N_{\mathrm{sym}}}$ of $N_{\mathrm{sym}}$ complex channel symbols through an $M$-point constellation $\mathcal{C}$. Let $\mathcal{H}(\cdot;\xi,L)$ represent the WDM fiber channel and
receiver digital signal processing (DSP), where $L$ is the link length and $\xi$ contains the launch power, amplifier-noise,
and neighboring-channel realizations. The semantic receiver $R_{\phi}$, with trainable parameters $\phi$, produces a reconstructed image $\hat{\mathbf{x}}$ and a class-logit vector $\hat{\mathbf{y}}$ according to
\begin{equation}
(\hat{\mathbf{x}},\hat{\mathbf{y}})
=
R_{\phi}\!\left(
\mathcal{H}\!\left(T_{\theta}(\mathbf{x};\mathcal{C});\xi,L\right);
\mathcal{C}
\right).
\end{equation}
Here, $\hat{\mathbf{y}} \in \mathbb{R}^{K}$ contains the classification logits, and the inferred class is $\hat{y} = \arg\max_{j} \hat{y}_{j}$
where $\hat{y}\in\{1,\ldots,K\}$.

The objective is to learn $\theta$, $\phi$, and, in the learned-constellation variant, $\mathcal{C}$, by minimizing a semantic loss $\mathcal{L}_{\mathrm{sem}}$ that jointly measures classification and reconstruction performance $
 \min_{\theta,\phi,\mathcal{C}}
 \mathbb{E}_{(\mathbf{x},y)\sim\mathcal{D},\boldsymbol{\xi}}
 \left[\mathcal{L}_{\mathrm{sem}}\right]
 \label{eq:problem}
$,
where the minimization over $\mathcal{C}$ applies only to the learned-constellation variant.

\subsection{Semantic Transmitter}\label{subsec:semantic-transmitter}
Fig.~\ref{fig:semcom_pipeline} illustrates the proposed system architecture. The forward path comprises the semantic transmitter, Manakov SSFM channel, soft demodulator, and semantic receiver. The two outputs define a common semantic loss that provides task-level feedback for optimizing the proposed system according to Eq.~\eqref{eq:problem}.

The semantic transmitter $T_{\theta}$ comprises a two-stage convolutional encoder $E_{\theta}$ followed by a linear logit head. The encoder extracts the semantic representation $\mathbf{z}$ from the source image $\mathbf{x}$. The logit head maps $\mathbf{z}$ to a logit vector $\ell_{\theta,n} \in \mathbb{R}^{M}$ for each of the
$N_{\mathrm{sym}}$ transmitted symbols, from which a softmax defines a categorical distribution
$p_{\theta,n}(\cdot \mid \mathbf{x})$ over the $M$ constellation points; the index $k_n$ for the $n$-th symbol is drawn from this distribution and mapped to the constellation point $c_{k_n}\in\mathcal{C}$.

The constellation $\mathcal{C}=\{c_m\}_{m=1}^{M}$ is initialized as Gray-labeled square QAM and normalized so that its points have unit average energy, i.e., $\frac{1}{M}\sum_{m=1}^{M}|c_m|^2=1$. In the learned variant, the in-phase and quadrature (I/Q) coordinates are optimized jointly with the semantic transmitter and semantic receiver; this constraint is imposed only at initialization, so the learned geometry is free to depart from unit average energy during training. In the fixed-constellation variant, $\mathcal{C}$ retains its initial QAM geometry and unit average energy.

Because the constellation energy is unconstrained after initialization, and because the symbol distribution $p_{\theta,n}(\cdot \mid \mathbf{x})$ is data-dependent and therefore non-uniform over $\mathcal{C}$, the empirical energy of a transmitted block is in general not unity. Each block is consequently normalized by its empirical symbol-energy
factor
\begin{equation}\label{eq:power-normalization}
a =
\left(
\frac{1}{N_{\mathrm{sym}}}
\sum_{n=1}^{N_{\mathrm{sym}}}
|c_{k_n}|^2
\right)^{1/2},
\end{equation}
and the transmitted symbols are defined as
$s_n=c_{k_n}/a$, such that $
\frac{1}{N_{\mathrm{sym}}}
\sum_{n=1}^{N_{\mathrm{sym}}}|s_n|^2=1
$.
This normalization prevents reductions in $\mathcal{L}_{\mathrm{sem}}$ from being achieved completely by increasing transmitted block energy, and separates launch-power control from constellation geometry.

\subsection{Optical Channel Model}\label{subsec:optical-channel}
The transmitted block in Eq.~\eqref{eq:power-normalization} is upsampled at $S_{\mathrm{ps}}$ samples per symbol and shaped by a unit-energy root-raised-cosine (RRC) filter with $N_{\mathrm{RRC}}$ taps and roll-off $\beta$. The waveform is scaled to the prescribed per-channel launch power $P_{\mathrm{tx}}$, which is controlled independently of the semantic representation. The channel of interest (CoI) is multiplexed with $N_{\mathrm{ch}}-1$ power-matched neighboring WDM channels, each carrying independent, identically distributed 16-QAM symbols. The neighbor modulation order is held at 16-QAM irrespective of the CoI order $M$, so that the inter-channel interference environment is identical across the $M$ sweep and any variation with $M$ is attributable to the CoI alone. Joint WDM propagation includes inter-channel nonlinear interference such as cross-phase modulation and four-wave mixing.

The composite optical field is propagated using the Manakov equation, solved numerically by SSFM~\cite{Marcuse1997Application}, with an adaptive step size based on a maximum nonlinear phase rotation criterion and bounded by $h$. The single-polarization launch is represented by setting the orthogonal component to zero, and polarization-mode dispersion is not modeled. Each fiber span is followed by an erbium-doped fiber amplifier (EDFA) that compensates span loss and adds ASE noise according to its noise figure $NF$, thereby including ASE noise, signal--signal and signal--noise Kerr interactions. At the receiver, standard coherent DSP applies full-link electronic dispersion compensation, CoI downconversion, matched filtering, symbol-rate downsampling, and per-block normalization to unit average power, which absorbs the launch scaling, span loss and amplifier gain into a single scalar~\cite{Savory2010Digital}.

Carrier-phase processing differs between the proposed system and the JPEG baselines. The proposed system applies no carrier-phase recovery and must accommodate the residual nonlinear phase rotation through its semantic receiver. In contrast, the JPEG baselines use ideal carrier-phase recovery with one maximum-likelihood phase estimate per block computed from the transmitted symbols. This choice provides the conventional baselines with an optimistic performance bound. The recovered symbol sequence $\{r_n\}$ is supplied to the soft demodulator. Numerical waveform, WDM, and link parameters are reported in Sec.~\ref{sec:experimental-protocol}.

\subsection{Semantic Receiver}\label{subsec:semantic-receiver}
The semantic receiver can be written as $R_\phi=G_\phi\circ D$, where $D(\cdot;\mathcal{C})$ is a soft-output max-log demodulator and $G_\phi$ is the neural semantic decoder. The normalized block is first rescaled by the transmit-side factor $a$, assumed known at the receiver, so that $D$ evaluates its likelihoods against the shared constellation $\mathcal{C}$ on the transmitted symbol scale. For each recovered symbol $r_n$, $D$ computes the log-likelihood ratio (LLR) for bit position $q$,
\begin{equation}
    \lambda_{n,q}=\min_{c\in\mathcal{C}_{q,0}}\lvert r_n-c\rvert^2
    -\min_{c\in\mathcal{C}_{q,1}}\lvert r_n-c\rvert^2,
\end{equation}
where $\mathcal{C}_{q,0}$ and $\mathcal{C}_{q,1}$ are the subsets of constellation points labeled 0 and 1 at bit position $q$, respectively. Stacking over all symbols and bit positions yields the LLR vector $\boldsymbol{\lambda}\in\mathbb{R}^{N_{\mathrm{sym}}\log_2 M}$, which is forwarded to the semantic decoder without hard decisions or channel decoding.

The neural semantic decoder $G_\phi$ maps $\boldsymbol{\lambda}$ to two task outputs through independent paths. The reconstruction head $\mathrm{Rec}_\phi(\boldsymbol{\lambda})$ projects
$\boldsymbol{\lambda}$ to a spatial feature volume via a fully connected layer, refines it through a residual block at the bottleneck resolution, and upsamples through two transposed-convolutional stages that mirror the spatial downsampling of $E_\theta$, yielding the flattened reconstructed image $\hat{\mathbf{x}}\in\mathbb{R}^{HW}$. In parallel, the classification head $C_\phi(\boldsymbol{\lambda})$ maps $\boldsymbol{\lambda}$ to class logits $\hat{\mathbf{y}}\in\mathbb{R}^{K}$ through two fully connected layers.
Both heads are trained jointly through the semantic loss $\mathcal{L}_{\mathrm{sem}}$.

\subsection{Training Procedure and Alternating Optimization}\label{subsec:training}
The channel model $\mathcal{H}$ is implemented using SSFM and is treated as non-differentiable because its Jacobian $\partial\mathcal{H}/\partial\mathbf{s}$ is unavailable to the learning algorithm. In addition, categorical sampling has no pathwise derivative with respect to $\theta$. 
Starting directly from random weights also creates a cold-start problem because the receiver initially observes an unstructured symbol mapping, which provides noisy reward signals for policy-gradient updates. Training therefore comprises differentiable pretraining over an AWGN surrogate followed by alternating receiver and transmitter optimization over the SSFM channel. The trainable variables are $\theta$, $\phi$, and $\mathcal{C}$ (the latter only in the learned-
constellation variant).

\subsubsection{Semantic Loss}\label{subsubsec:semantic-loss}
The semantic loss combines cross-entropy (CE) for classification, mean-squared error (MSE) for pixel fidelity, and the structural similarity index measure (SSIM)~\cite{Wang2004ImageQuality}:
\begin{align}
 \mathcal{L}_{\mathrm{sem}}
 ={}&\lambda_{\mathrm{MSE}}
 \operatorname{MSE}(\hat{\mathbf{x}},\mathbf{x})
 +\lambda_{\mathrm{SSIM}}
 \left[1-\operatorname{SSIM}(\hat{\mathbf{x}},\mathbf{x})\right]
 \nonumber\\
 &+\lambda_{\mathrm{CE}}
 \operatorname{CE}(\hat{\mathbf{y}},y),
 \label{eq:semantic-loss}
\end{align}
where CE promotes correct classification, MSE penalizes pixel-wise reconstruction error, and SSIM preserves local image structure that MSE alone may smooth. 

\subsubsection{Phase 1: Differentiable Pretraining}\label{subsubsection:pretraining}
Pretraining 
replaces SSFM with differentiable AWGN and categorical sampling with a hard straight-through Gumbel Softmax estimator~\cite{Jang2017Categorical}. The forward pass selects a constellation point, while the backward pass differentiates through its continuous relaxation to initialize $\theta$, $\phi$, as well as $\mathcal{C}$ in the learned-constellation variant, before black-box optimization.

\subsubsection{Phase 2: Alternating Black-Box Optimization}\label{subsubsec:alternating-optimization}
After pretraining, the AWGN surrogate is discarded and alternating black-box optimization over the SSFM channel runs for $N_{\mathrm{alt}}$ outer iterations. Within each outer iteration, the semantic receiver is updated for $N_{\mathrm{RX}}$ gradient steps, followed by $N_{\mathrm{TX}}$ policy-gradient steps for the semantic transmitter.

\textbf{Receiver update.} The semantic transmitter parameters $\theta$ are frozen, and symbols are selected deterministically according to $k_n=\arg\max_m p_{\theta,n}(m\mid\mathbf{x})$. The SSFM channel propagates these symbols, after which gradients flow through the soft demodulator and semantic receiver to update $\phi$. In the learned-constellation variant, $\mathcal{C}$ is updated through the differentiable LLR metric; no gradient is propagated through the SSFM channel or into $\theta$.

\textbf{Transmitter update.} The semantic receiver parameters $\phi$ and constellation $\mathcal{C}$ are frozen. Symbol indices $k_n$ are sampled from $p_{\theta,n}(\cdot\mid\mathbf{x})$, and the frozen receiver assigns the reward $R=-\mathcal{L}_{\mathrm{sem}}$ after SSFM propagation. Defining the transmitter objective as $
J(\theta) = \mathbb{E}[R]$,
the semantic transmitter parameters $\theta$ are updated using the score-function estimator~\cite{Williams1992Simple}:
\begin{equation}
 \nabla_{\theta}J
 =\mathbb{E}\!\left[
 (R-b)\nabla_{\theta}
 \sum_{n=1}^{N_{\mathrm{sym}}}
 \log p_{\theta,n}(k_n\mid\mathbf{x})
 \right],
 \label{eq:reinforce}
\end{equation}
where $b$ is an exponential-moving-average reward baseline updated across mini-batches and outer iterations. Rewards above or below $b$ respectively increase or decrease the probabilities of the selected symbols, reducing estimator variance without gradients through categorical sampling or the channel. For efficient per-sample rewards, the structural term is approximated by $1-\exp(-\operatorname{MSE})$, while true SSIM is retained for receiver training and evaluation. Freezing $\mathcal{C}$ prevents simultaneous changes to the policy and discrete action geometry.

\begin{figure*}[t]
\centering
\lengthkey\\[0.15em]

\begin{subfigure}[t]{0.49\textwidth}
\centering
\begin{tikzpicture}
\begin{axis}[pwraxis,
  ylabel={Classification accuracy [\%]},
  ymin=96.0, ymax=99.7, ytick={96,97,98,99},
  legend pos=south west,
]
  \addplot[c150,mSemL,forget plot] coordinates {(-9,99.22)(-6,99.22)(-3,99.25)(-1,99.24)(0,99.25)(1,99.24)(3,99.07)};
  \addplot[c200,mSemL,forget plot] coordinates {(-9,99.26)(-6,99.28)(-3,99.29)(-1,99.34)(0,99.31)(1,99.27)(3,98.95)};
  \addplot[c300,mSemL,forget plot] coordinates {(-9,99.12)(-6,99.15)(-3,99.15)(-1,99.16)(0,99.19)(1,99.21)(3,98.78)};
  \addplot[c400,mSemL,forget plot] coordinates {(-9,99.22)(-6,99.29)(-3,99.24)(-1,99.22)(0,99.17)(1,99.19)(3,98.83)};
  \addplot[c500,mSemL,forget plot] coordinates {(-9,99.08)(-6,99.12)(-3,99.07)(-1,99.09)(0,99.06)(1,99.06)(3,98.54)};
  \addplot[c600,mSemL,forget plot] coordinates {(-9,99.08)(-6,99.07)(-3,99.15)(-1,99.14)(0,99.07)(1,99.14)(3,98.38)};
  \addplot[c700,mSemL,forget plot] coordinates {(-9,98.88)(-6,98.99)(-3,99.12)(-1,99.16)(0,99.10)(1,99.07)(3,98.02)};
  \addplot[c800,mSemL,forget plot] coordinates {(-9,98.73)(-6,98.88)(-3,99.18)(-1,99.26)(0,99.24)(1,99.17)(3,98.17)};

  \addplot[c150,mSemF,forget plot] coordinates {(-9,99.15)(-6,99.17)(-3,99.23)(-1,99.23)(0,99.21)(1,99.20)(3,98.95)};
  \addplot[c200,mSemF,forget plot] coordinates {(-9,99.17)(-6,99.16)(-3,99.16)(-1,99.19)(0,99.17)(1,99.19)(3,98.90)};
  \addplot[c300,mSemF,forget plot] coordinates {(-9,99.18)(-6,99.19)(-3,99.19)(-1,99.16)(0,99.16)(1,99.13)(3,98.67)};
  \addplot[c400,mSemF,forget plot] coordinates {(-9,99.02)(-6,99.01)(-3,99.06)(-1,99.06)(0,99.13)(1,99.09)(3,98.76)};
  \addplot[c500,mSemF,forget plot] coordinates {(-9,98.87)(-6,98.94)(-3,99.03)(-1,99.07)(0,99.05)(1,99.08)(3,98.31)};
  \addplot[c600,mSemF,forget plot] coordinates {(-9,98.91)(-6,99.04)(-3,99.15)(-1,99.20)(0,99.20)(1,99.20)(3,98.29)};
  \addplot[c700,mSemF,forget plot] coordinates {(-9,98.89)(-6,99.02)(-3,99.08)(-1,99.06)(0,98.98)(1,99.05)(3,97.79)};
  \addplot[c800,mSemF,forget plot] coordinates {(-9,97.79)(-6,98.80)(-3,99.13)(-1,99.18)(0,99.16)(1,99.12)(3,96.82)};

  \addlegendimage{black,mSemL}\addlegendentry{Proposed system (learned constellation)}
  \addlegendimage{black,mSemF}\addlegendentry{Proposed system (fixed constellation)}
\end{axis}
\end{tikzpicture}
\caption{Classification accuracy of the proposed system}
\label{fig:acc-semantic}
\end{subfigure}
\hfill
\begin{subfigure}[t]{0.49\textwidth}
\centering
\begin{tikzpicture}
\begin{axis}[pwraxis,
  ylabel={Classification accuracy [\%]},
  ymin=-4, ymax=104, ytick={0,20,40,60,80,100},
  legend pos=south west,
]
  \fill[gray!14] (axis cs:-9.8,96.82) rectangle (axis cs:3.8,99.34);
  \draw[gray!55, line width=0.3pt, densely dotted]
        (axis cs:-9.8,96.82) -- (axis cs:3.8,96.82);
  \draw[gray!55, line width=0.3pt, densely dotted]
        (axis cs:-9.8,99.34) -- (axis cs:3.8,99.34);
  \node[font=\scriptsize, anchor=south, align=center, gray!35!black]
        (envlab) at (axis cs:-4.5,88.0) {proposed, all $L$};
  \draw[gray!55, line width=0.3pt, ->] (envlab.north) -- (axis cs:-4.5,96.22);

  \addplot[c150,mLDPC,forget plot] coordinates {(-9,85.55)(-6,85.55)(-3,85.55)(-1,85.55)(0,85.55)(1,85.55)(3,85.54)};
  \addplot[c200,mLDPC,forget plot] coordinates {(-9,85.55)(-6,85.55)(-3,85.55)(-1,85.55)(0,85.55)(1,85.54)(3,85.34)};
  \addplot[c300,mLDPC,forget plot] coordinates {(-9,85.57)(-6,85.57)(-3,85.57)(-1,85.56)(0,85.54)(1,85.32)(3,84.08)};
  \addplot[c400,mLDPC,forget plot] coordinates {(-9,85.55)(-6,85.55)(-3,85.55)(-1,85.40)(0,85.01)(1,84.75)(3,82.59)};
  \addplot[c500,mLDPC,forget plot] coordinates {(-9,85.57)(-6,85.57)(-3,85.57)(-1,85.51)(0,85.15)(1,85.52)(3,84.36)};
  \addplot[c600,mLDPC,forget plot] coordinates {(-9,85.57)(-6,85.57)(-3,85.57)(-1,85.22)(0,85.39)(1,85.49)(3,73.39)};
  \addplot[c700,mLDPC,forget plot] coordinates {(-9,85.57)(-6,85.57)(-3,85.56)(-1,84.90)(0,85.54)(1,85.32)(3,28.78)};
  \addplot[c800,mLDPC,forget plot] coordinates {(-9,85.57)(-6,85.57)(-3,85.52)(-1,85.54)(0,85.48)(1,85.03)(3,3.06)};

  \addplot[c150,mBare,forget plot] coordinates {(-9,85.55)(-6,85.55)(-3,85.55)(-1,85.55)(0,85.55)(1,85.50)(3,71.17)};
  \addplot[c200,mBare,forget plot] coordinates {(-9,85.55)(-6,85.55)(-3,85.55)(-1,85.55)(0,85.54)(1,84.97)(3,32.37)};
  \addplot[c300,mBare,forget plot] coordinates {(-9,85.55)(-6,85.55)(-3,85.55)(-1,85.55)(0,85.09)(1,75.19)(3,0.57)};
  \addplot[c400,mBare,forget plot] coordinates {(-9,85.55)(-6,85.55)(-3,85.55)(-1,84.93)(0,73.84)(1,23.33)(3,0.00)};
  \addplot[c500,mBare,forget plot] coordinates {(-9,85.55)(-6,85.55)(-3,85.50)(-1,78.12)(0,30.89)(1,0.52)(3,0.00)};
  \addplot[c600,mBare,forget plot] coordinates {(-9,85.50)(-6,85.55)(-3,84.97)(-1,46.55)(0,1.72)(1,0.01)(3,0.00)};
  \addplot[c700,mBare,forget plot] coordinates {(-9,85.20)(-6,85.54)(-3,82.53)(-1,7.92)(0,0.00)(1,0.00)(3,0.00)};
  \addplot[c800,mBare,forget plot] coordinates {(-9,84.70)(-6,85.29)(-3,68.08)(-1,0.21)(0,0.00)(1,0.00)(3,0.00)};

  \addlegendimage{black,mLDPC}\addlegendentry{LDPC-coded JPEG}
  \addlegendimage{black,mBare}\addlegendentry{Uncoded JPEG}
\end{axis}
\end{tikzpicture}
\caption{JPEG-baseline classification accuracy.}
\label{fig:acc-baseline}
\end{subfigure}

\vspace{0.15em}

\begin{subfigure}[t]{0.49\textwidth}
\centering
\begin{tikzpicture}
\begin{axis}[pwraxis,
  ylabel={SSIM},
  ymin=0.62, ymax=0.87, ytick={0.65,0.70,0.75,0.80,0.85},
  legend pos=south west,
]
  \addplot[c150,mSemL,forget plot] coordinates {(-9,0.8414)(-6,0.8424)(-3,0.8450)(-1,0.8472)(0,0.8475)(1,0.8461)(3,0.7993)};
  \addplot[c200,mSemL,forget plot] coordinates {(-9,0.8426)(-6,0.8452)(-3,0.8499)(-1,0.8513)(0,0.8509)(1,0.8486)(3,0.7993)};
  \addplot[c300,mSemL,forget plot] coordinates {(-9,0.8264)(-6,0.8325)(-3,0.8407)(-1,0.8431)(0,0.8424)(1,0.8388)(3,0.7982)};
  \addplot[c400,mSemL,forget plot] coordinates {(-9,0.8267)(-6,0.8329)(-3,0.8416)(-1,0.8427)(0,0.8407)(1,0.8360)(3,0.7911)};
  \addplot[c500,mSemL,forget plot] coordinates {(-9,0.8083)(-6,0.8216)(-3,0.8370)(-1,0.8407)(0,0.8390)(1,0.8351)(3,0.7674)};
  \addplot[c600,mSemL,forget plot] coordinates {(-9,0.7708)(-6,0.7932)(-3,0.8249)(-1,0.8301)(0,0.8289)(1,0.8256)(3,0.7344)};
  \addplot[c700,mSemL,forget plot] coordinates {(-9,0.7537)(-6,0.7883)(-3,0.8271)(-1,0.8351)(0,0.8348)(1,0.8289)(3,0.7321)};
  \addplot[c800,mSemL,forget plot] coordinates {(-9,0.7018)(-6,0.7537)(-3,0.8218)(-1,0.8294)(0,0.8288)(1,0.8207)(3,0.7241)};

  \addplot[c150,mSemF,forget plot] coordinates {(-9,0.8418)(-6,0.8439)(-3,0.8466)(-1,0.8480)(0,0.8481)(1,0.8470)(3,0.8281)};
  \addplot[c200,mSemF,forget plot] coordinates {(-9,0.8376)(-6,0.8406)(-3,0.8450)(-1,0.8472)(0,0.8475)(1,0.8456)(3,0.8055)};
  \addplot[c300,mSemF,forget plot] coordinates {(-9,0.8278)(-6,0.8344)(-3,0.8415)(-1,0.8424)(0,0.8391)(1,0.8404)(3,0.7747)};
  \addplot[c400,mSemF,forget plot] coordinates {(-9,0.8176)(-6,0.8282)(-3,0.8406)(-1,0.8405)(0,0.8414)(1,0.8350)(3,0.7874)};
  \addplot[c500,mSemF,forget plot] coordinates {(-9,0.7825)(-6,0.8071)(-3,0.8365)(-1,0.8407)(0,0.8325)(1,0.8233)(3,0.7513)};
  \addplot[c600,mSemF,forget plot] coordinates {(-9,0.7628)(-6,0.7949)(-3,0.8311)(-1,0.8345)(0,0.8289)(1,0.8284)(3,0.7206)};
  \addplot[c700,mSemF,forget plot] coordinates {(-9,0.7338)(-6,0.7796)(-3,0.8257)(-1,0.8260)(0,0.8238)(1,0.8210)(3,0.7009)};
  \addplot[c800,mSemF,forget plot] coordinates {(-9,0.7040)(-6,0.7639)(-3,0.8192)(-1,0.8242)(0,0.8266)(1,0.8174)(3,0.6624)};

  \addlegendimage{black,mSemL}\addlegendentry{Proposed system (learned constellation)}
  \addlegendimage{black,mSemF}\addlegendentry{Proposed system (fixed constellation)}
\end{axis}
\end{tikzpicture}
\caption{SSIM of the proposed system.}
\label{fig:ssim-semantic}
\end{subfigure}
\hfill
\begin{subfigure}[t]{0.49\textwidth}
\centering
\begin{tikzpicture}
\begin{axis}[pwraxis,
  ylabel={SSIM},
  ymin=-0.04, ymax=1.0, ytick={0,0.2,0.4,0.6,0.8,1.0},
  legend pos=south west,
]
  \fill[gray!14] (axis cs:-9.8,0.6624) rectangle (axis cs:3.8,0.8513);
  \draw[gray!55, line width=0.3pt, densely dotted]
        (axis cs:-9.8,0.6624) -- (axis cs:3.8,0.6624);
  \draw[gray!55, line width=0.3pt, densely dotted]
        (axis cs:-9.8,0.8513) -- (axis cs:3.8,0.8513);
  \node[font=\scriptsize, anchor=south, align=center, gray!35!black]
        (envlab) at (axis cs:-4.5,0.4) {proposed, all $L$};
  \draw[gray!55, line width=0.3pt, ->] (envlab.north) -- (axis cs:-4.5,0.6424);

  \addplot[c150,mLDPC,forget plot] coordinates {(-9,0.8693)(-6,0.8693)(-3,0.8693)(-1,0.8693)(0,0.8693)(1,0.8693)(3,0.8693)};
  \addplot[c200,mLDPC,forget plot] coordinates {(-9,0.8693)(-6,0.8693)(-3,0.8693)(-1,0.8693)(0,0.8693)(1,0.8693)(3,0.8675)};
  \addplot[c300,mLDPC,forget plot] coordinates {(-9,0.8693)(-6,0.8693)(-3,0.8693)(-1,0.8693)(0,0.8690)(1,0.8665)(3,0.8538)};
  \addplot[c400,mLDPC,forget plot] coordinates {(-9,0.8693)(-6,0.8693)(-3,0.8693)(-1,0.8684)(0,0.8642)(1,0.8619)(3,0.8386)};
  \addplot[c500,mLDPC,forget plot] coordinates {(-9,0.8693)(-6,0.8693)(-3,0.8693)(-1,0.8687)(0,0.8654)(1,0.8689)(3,0.8572)};
  \addplot[c600,mLDPC,forget plot] coordinates {(-9,0.8693)(-6,0.8693)(-3,0.8693)(-1,0.8652)(0,0.8672)(1,0.8685)(3,0.7446)};
  \addplot[c700,mLDPC,forget plot] coordinates {(-9,0.8693)(-6,0.8693)(-3,0.8689)(-1,0.8632)(0,0.8691)(1,0.8668)(3,0.2916)};
  \addplot[c800,mLDPC,forget plot] coordinates {(-9,0.8693)(-6,0.8693)(-3,0.8688)(-1,0.8691)(0,0.8684)(1,0.8637)(3,0.0321)};

  \addplot[c150,mBare,forget plot] coordinates {(-9,0.8693)(-6,0.8693)(-3,0.8693)(-1,0.8693)(0,0.8693)(1,0.8687)(3,0.7234)};
  \addplot[c200,mBare,forget plot] coordinates {(-9,0.8693)(-6,0.8693)(-3,0.8693)(-1,0.8693)(0,0.8693)(1,0.8637)(3,0.3431)};
  \addplot[c300,mBare,forget plot] coordinates {(-9,0.8693)(-6,0.8693)(-3,0.8693)(-1,0.8693)(0,0.8652)(1,0.7709)(3,0.0047)};
  \addplot[c400,mBare,forget plot] coordinates {(-9,0.8693)(-6,0.8693)(-3,0.8693)(-1,0.8647)(0,0.7596)(1,0.2452)(3,0.0000)};
  \addplot[c500,mBare,forget plot] coordinates {(-9,0.8693)(-6,0.8693)(-3,0.8690)(-1,0.8063)(0,0.3283)(1,0.0058)(3,0.0000)};
  \addplot[c600,mBare,forget plot] coordinates {(-9,0.8688)(-6,0.8693)(-3,0.8671)(-1,0.4863)(0,0.0185)(1,0.0001)(3,0.0000)};
  \addplot[c700,mBare,forget plot] coordinates {(-9,0.8660)(-6,0.8692)(-3,0.8482)(-1,0.0850)(0,0.0001)(1,0.0000)(3,0.0000)};
  \addplot[c800,mBare,forget plot] coordinates {(-9,0.8613)(-6,0.8670)(-3,0.7096)(-1,0.0023)(0,0.0000)(1,0.0000)(3,0.0000)};

  \addlegendimage{black,mLDPC}\addlegendentry{LDPC-coded JPEG}
  \addlegendimage{black,mBare}\addlegendentry{Uncoded JPEG}
\end{axis}
\end{tikzpicture}
\caption{JPEG-baseline SSIM.}
\label{fig:ssim-baseline}
\end{subfigure}
\caption{\footnotesize Classification accuracy (top) and SSIM (bottom) versus launch power at $M=16$. Color indicates link length. The proposed system is shown on the left and the JPEG baselines on the right, where shading marks the performance range of the proposed system.}
\vspace{-1.0em}
\label{fig:perf-vs-power}
\label{fig:acc-vs-power}
\label{fig:ssim-vs-power}
\end{figure*}
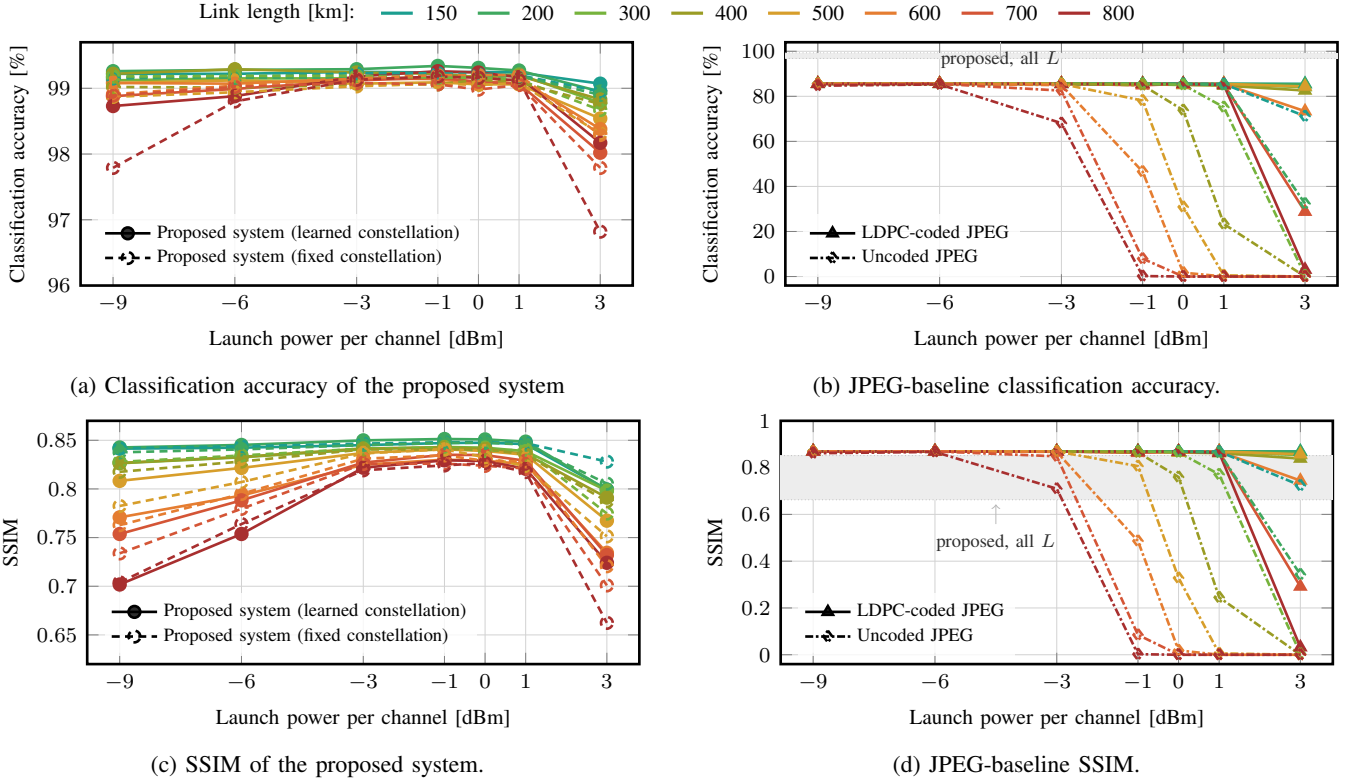

\section{Experimental Settings} \label{sec:experimental-protocol}

\begin{table}[b]
\vspace{-0.3cm}
\centering
\caption{Optical waveform and fiber parameters.}
\label{tab:optical-parameters}
\footnotesize
\renewcommand{\arraystretch}{0.94}
\begin{tabular}{@{}p{0.16\columnwidth}p{0.39\columnwidth}p{0.33\columnwidth}@{}}
\hline
Symbol & Quantity & Value \\
\hline
$R_{\mathrm{sym}}$ & Symbol rate & 32~GBaud \\
$S_{\mathrm{ps}}$ & Samples per symbol & 16 \\
$N_{\mathrm{RRC}}$ & RRC filter length & 1024 taps \\
$\beta$ & RRC roll-off factor & 0.01 \\
$N_{\mathrm{ch}}$ & Number of WDM channels & 11 \\
$\Delta f$ & WDM channel spacing & 37.5~GHz \\
$k_{\mathrm{CoI}}$ & CoI position & 0 (center)\\
$M_{\mathrm{nbr}}$ & Neighbor modulation order & 16-QAM \\
$L$ & Link length & [150, 200, 300, 400, 500, 600, 700, 800]~km \\
$L_{\mathrm{span}}$ & Span length & 50~km \\
$h$ & Nominal SSFM step & 0.5~km, phase-adaptive \\
$\alpha$ & Attenuation coefficient & 0.2~dB/km \\
$D$ & Dispersion parameter & 16~ps/(nm$\cdot$km) \\
$\gamma$ & Kerr coefficient & 1.3~W$^{-1}$km$^{-1}$ \\
$F_c$ & Carrier frequency & 193.1~THz \\
$NF$ & EDFA noise figure & 4.5~dB \\
\hline
\end{tabular}
\end{table}

\textbf{Data.} Experiments use the MNIST dataset with $H=W=28$ and $K=10$, including 60,000 training images and 10,000 test images~\cite{lecun1998gradient}. Pixel intensities are normalized to $[0,1]$, and the same test set is used across all experiments for evaluation.

\textbf{Model variants.} Each image is transmitted using a fixed block of $N_{\mathrm{sym}}=128$ symbols, while the QAM order varies over 
$M\in\{16,64,256\}$. For both learned and fixed constellations the CoI is placed at the center of the WDM grid.

\textbf{Neural architecture.} Across both system variants, the embedding and decoder hidden dimensions are fixed at 32 and 256, respectively, and training uses a batch size of 32. 

\textbf{Optical channel configuration.} Tab.~\ref{tab:optical-parameters} lists the waveform, WDM, and fiber parameters. Before launch-power scaling, each waveform generated by the proposed system is normalized to unit average power; hence, the reported launch power denotes the per-channel average power at the fiber input. Waveform generation, filtering, and propagation are implemented using OptiCommPy~\cite{opticommpy}.

\textbf{Training configuration.} The semantic-loss weights are $\lambda_{\mathrm{MSE}}=1$, $\lambda_{\mathrm{SSIM}}=0.5$, and $\lambda_{\mathrm{CE}}=1$. AWGN pretraining runs for 10 epochs at a signal-to-noise ratio (SNR) of 20~dB. Then, Manakov-SSFM fine-tuning uses $N_{\mathrm{alt}}=100$ outer iterations, each with $N_{\mathrm{RX}}=50$ receiver gradient and $N_{\mathrm{TX}}=50$ transmitter policy-gradient steps per iteration, with the launch power sampled uniformly from $[-3,+3]$~dBm. A separate model is trained for each combination of link length,
modulation order, and constellation variant, resulting in $48$ systems
with identical architectures, hyperparameters, and training budgets.


\textbf{Baselines.} Two baselines apply JPEG compression with quality factor $Q=25$, which controls the tradeoff between compression and reconstruction fidelity. The \emph{uncoded JPEG} baseline transmits the compressed bitstream directly, whereas the \emph{LDPC-coded JPEG} baseline applies rate-$1/2$ LDPC coding~\cite{Richardson2001Efficient}. After channel demodulation and, for the coded baseline, LDPC decoding, successfully recovered JPEG images are classified by a separately trained ResNet-18~\cite{He2016Deep}, an 18-layer residual CNN followed by a 10-class linear head, trained on uncompressed samples. The same classifier is used for both baselines and all channel conditions. A channel-free control also classified the $Q=25$ JPEG images after compression and decompression without optical propagation, thereby isolating the source-compression effect. If JPEG decoding fails, the frame is counted as a classification error and assigned zero SSIM because no image is available for classification. Both baselines use the same optical channel as the proposed system and are compared at equal per-channel average launch power.


\textbf{Evaluation.} Performance is evaluated at launch powers of $-9,-6,-3,-1,0,1$, and $3$~dBm. Classification accuracy is used as task metric, while SSIM quantifies reconstruction quality.

\begin{figure*}[t]
\centering
\modkey\par\vspace{0.10em}
\schemekey\par\vspace{0.20em}
 
\begin{subfigure}[t]{0.49\textwidth}
\centering
\begin{tikzpicture}
\begin{axis}[reachaxis,
  ylabel={Classification accuracy [\%]},
  ymin=-5, ymax=104, ytick={0,20,40,60,80,100},
]
  \addplot[cM16,mSemL] coordinates {(150,99.25)(200,99.31)(300,99.19)(400,99.17)(500,99.06)(600,99.07)(700,99.10)(800,99.24)};
  \addplot[cM64,mSemL] coordinates {(150,99.21)(200,99.11)(300,99.06)(400,99.23)(500,99.12)(600,99.12)(700,98.96)(800,99.04)};
  \addplot[cM256,mSemL] coordinates {(150,99.20)(200,99.02)(300,99.19)(400,98.92)(500,99.11)(600,99.00)(700,99.17)(800,99.04)};
  \addplot[cM16,mSemF] coordinates {(150,99.21)(200,99.17)(300,99.16)(400,99.13)(500,99.05)(600,99.20)(700,98.98)(800,99.16)};
  \addplot[cM64,mSemF] coordinates {(150,99.23)(200,99.27)(300,99.15)(400,99.20)(500,99.05)(600,99.08)(700,99.02)(800,99.20)};
  \addplot[cM256,mSemF] coordinates {(150,99.21)(200,99.25)(300,99.17)(400,99.08)(500,99.07)(600,99.18)(700,99.15)(800,99.11)};
  \addplot[cM16,mLDPC] coordinates {(150,85.55)(200,85.55)(300,85.54)(400,85.01)(500,85.15)(600,85.39)(700,85.54)(800,85.48)};
  \addplot[cM64,mLDPC] coordinates {(150,85.55)(200,85.55)(300,85.50)(400,85.43)(500,82.95)(600,63.46)(700,30.46)(800,11.02)};
  \addplot[cM256,mLDPC] coordinates {(150,85.10)(200,84.15)(300,59.22)(400,10.63)(500,0.67)(600,0.00)(700,0.00)(800,0.00)};
  \addplot[cM16,mBare] coordinates {(150,85.55)(200,85.54)(300,85.09)(400,73.84)(500,30.89)(600,1.72)(700,0.00)(800,0.00)};
  \addplot[cM64,mBare] coordinates {(150,0.15)(200,0.00)(300,0.00)(400,0.00)(500,0.00)(600,0.00)(700,0.00)(800,0.00)};
  \addplot[cM256,mBare] coordinates {(150,0.00)(200,0.00)(300,0.00)(400,0.00)(500,0.00)(600,0.00)(700,0.00)(800,0.00)};
\end{axis}
\end{tikzpicture}
\caption{Accuracy versus link length at $0$~dBm.}
\label{fig:reach-acc}
\end{subfigure}
\hfill
\begin{subfigure}[t]{0.49\textwidth}
\centering
\begin{tikzpicture}
\begin{axis}[reachaxis,
  ylabel={SSIM},
  ymin=-0.04, ymax=1.0, ytick={0,0.2,0.4,0.6,0.8,1.0},
]
  \addplot[cM16,mSemL] coordinates {(150,0.8475)(200,0.8509)(300,0.8424)(400,0.8407)(500,0.8390)(600,0.8289)(700,0.8348)(800,0.8288)};
  \addplot[cM64,mSemL] coordinates {(150,0.8342)(200,0.8362)(300,0.8260)(400,0.8278)(500,0.8281)(600,0.8207)(700,0.8202)(800,0.8164)};
  \addplot[cM256,mSemL] coordinates {(150,0.8325)(200,0.8296)(300,0.8213)(400,0.8186)(500,0.8208)(600,0.8174)(700,0.8088)(800,0.8084)};
  \addplot[cM16,mSemF] coordinates {(150,0.8481)(200,0.8475)(300,0.8391)(400,0.8414)(500,0.8325)(600,0.8289)(700,0.8238)(800,0.8266)};
  \addplot[cM64,mSemF] coordinates {(150,0.8471)(200,0.8466)(300,0.8372)(400,0.8352)(500,0.8299)(600,0.8304)(700,0.8277)(800,0.8252)};
  \addplot[cM256,mSemF] coordinates {(150,0.8415)(200,0.8409)(300,0.8371)(400,0.8266)(500,0.8266)(600,0.8197)(700,0.8205)(800,0.8179)};
  \addplot[cM16,mLDPC] coordinates {(150,0.8693)(200,0.8693)(300,0.8690)(400,0.8642)(500,0.8654)(600,0.8672)(700,0.8691)(800,0.8684)};
  \addplot[cM64,mLDPC] coordinates {(150,0.8693)(200,0.8693)(300,0.8688)(400,0.8681)(500,0.8430)(600,0.6504)(700,0.3187)(800,0.1181)};
  \addplot[cM256,mLDPC] coordinates {(150,0.8655)(200,0.8572)(300,0.6134)(400,0.1112)(500,0.0073)(600,0.0000)(700,0.0000)(800,0.0000)};
  \addplot[cM16,mBare] coordinates {(150,0.8693)(200,0.8693)(300,0.8652)(400,0.7596)(500,0.3283)(600,0.0185)(700,0.0001)(800,0.0000)};
  \addplot[cM64,mBare] coordinates {(150,0.0014)(200,0.0000)(300,0.0000)(400,0.0000)(500,0.0000)(600,0.0000)(700,0.0000)(800,0.0000)};
  \addplot[cM256,mBare] coordinates {(150,0.0000)(200,0.0000)(300,0.0000)(400,0.0000)(500,0.0000)(600,0.0000)(700,0.0000)(800,0.0000)};
\end{axis}
\end{tikzpicture}
\caption{SSIM versus link length at $0$~dBm.}
\label{fig:reach-ssim}
\end{subfigure}
 
\caption{\footnotesize Classification accuracy (a) and SSIM (b) versus link length at nominal launch power of $0$~dBm. Color indicates QAM order $M\in\{16,64,256\}$, while marker and line style indicate the system variant.}
\vspace{-1.5em}
\label{fig:reach}
\end{figure*}
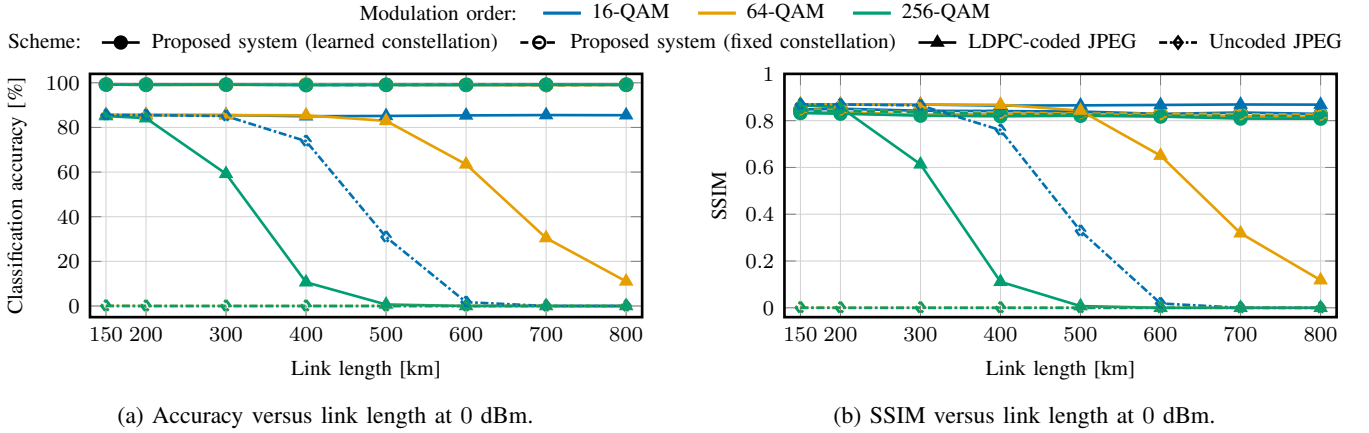

\section{Results and Discussion}\label{sec:results}
\subsection{Robustness to launch power and reach}
Fig.~\ref{fig:perf-vs-power} compares the classification accuracy and the SSIM of the proposed semantic system and the two baselines using 16-QAM transmission, across launch powers and link lengths. Fig.~\ref{fig:acc-semantic} shows that the accuracy of the semantic transmission system remains stable from $-9$ to $+1$~dBm, varying by at most $0.6$ percentage points at any link length, with a loss below one point at $+3$~dBm. Since the system is trained with launch power drawn from $[-3,+3]$~dBm, the stability at $-9$ and $-6$~dBm is an out-of-distribution result rather than a fitted one. In contrast, Fig.~\ref{fig:acc-baseline} shows the accuracy trend of the JPEG baselines. When JPEG decoding succeeds, classification accuracy saturates at approximately $85\%$. The channel-free JPEG control yields the same accuracy ceiling, confirming that the plateau is imposed by the lossy JPEG source-compression setting, rather than by the optical channel. In the LDPC-coded baseline, FEC can recover the compressed bitstream but cannot restore information discarded by the source coder. At $+3$~dBm, the accuracy of the \emph{LDPC-coded JPEG} baseline falls to $73.4\%$, $28.8\%$, and $3.1\%$ at $600$, $700$, and $800$~km, respectively, while the \emph{uncoded JPEG} baseline reaches zero by $-1$~dBm at $800$~km, its collapse migrating to lower power as the link lengthens. This behavior reflects the sensitivity of the entropy-coded JPEG stream to residual bit errors, where a single uncorrected error can invalidate the payload.

The SSIM trends reported in Fig.~\ref{fig:ssim-semantic} reveal an operating-power optimum between $-1$ and $0$~dBm, balancing low-power ASE noise and high-power Kerr nonlinearity. However, the SSIM of the semantic communication system stays above $0.66$ throughout the sweep. By contrast, Fig.~\ref{fig:ssim-baseline} shows an SSIM near $0.869$ when JPEG decoding succeeds and near zero when it fails. JPEG therefore provides higher reconstruction fidelity under reliable decoding, whereas the proposed system degrades gradually beyond the baseline failure point. 
The learned and fixed constellations perform almost identically.

\subsection{Reach at nominal launch power}
Fig.~\ref{fig:reach-acc} shows that at $0$~dBm the proposed system maintains $98.92$--$99.31\%$ accuracy from $150$ to $800$~km at every QAM order, with at most a $0.23$ percentage point difference between the learned and fixed constellations. 
In contrast, the \emph{uncoded JPEG} baseline falls from $85.09\%$ at $300$~km to $1.72\%$ at $600$~km with $16$-QAM and is nearly unusable at the shortest link length with higher QAM orders. The \emph{LDPC-coded JPEG} baseline preserves approximately $85\%$ accuracy through $800$~km with $16$-QAM. This ceiling holds only while decoding succeeds, as accuracy still falls to $11.02\%$ at $800$~km with $64$-QAM and to $0.67\%$ at $500$~km with $256$-QAM, because a denser constellation reduces the noise margin at fixed launch power. FEC therefore extends reach but does not remove the dependence on QAM order.

Fig.~\ref{fig:reach-ssim} shows a similar pattern in terms of reconstruction quality. The SSIM of the semantic communication system remains between $0.808$ and $0.851$, whereas the JPEG baselines' SSIM drops sharply once decoding becomes unreliable. 

\subsection{Transmission cost}
\begin{table}[b]
\centering
\caption{Transmission cost per image at 16/64/256-QAM.}\label{tab:cost}
\begin{tabular}{lcc}
\hline
Scheme & Symbols/image & Transmission time [ns] \\
\hline
Semantic system & 128/128/128 & 4.0/4.0/4.0 \\
Uncoded JPEG & 233/155/117 & 7.3/4.9/3.7 \\
LDPC-coded JPEG & 466/311/233 & 14.6/9.7/7.3 \\
\hline
\end{tabular}
\end{table}

At $R_{\mathrm{sym}}=32$~GBaud, each symbol has duration $T_{\mathrm{sym}}=1/R_{\mathrm{sym}}=31.25$~ps and carries $\log_2 M$ bits. The transmission time per image, $\tau=N_{\mathrm{sym}}/R_{\mathrm{sym}}$, therefore depends on symbol count but not on the QAM order. The semantic communication system transmits $N_{\mathrm{sym}}=128$ symbols per image, fixing its transmission time at $4.0$~ns. The JPEG baselines instead transmit fixed entropy-coded payloads, so their symbol counts and transmission times decrease as $M$ grows. Table~\ref{tab:cost} shows that the semantic communication system uses fewer symbols and less transmission time than the \emph{LDPC-coded JPEG} baseline at every QAM order: $128$ versus $466/311/233$ symbols and $4.0$~ns versus $14.6/9.7/7.3$~ns. The \emph{uncoded JPEG} baseline becomes slightly cheaper only at $256$-QAM, where it is unreliable beyond short link lengths. Thus, the proposed system achieves lower transmission cost than the LDPC-coded JPEG baseline while maintaining a constant transmission time across modulation orders.

\section{Conclusion}\label{sec:conclusion}
This work presented an end-to-end optical semantic communication
system for joint image classification and reconstruction over a
nonlinear WDM fiber link. By optimizing task performance directly over
a non-differentiable Manakov-SSFM channel, the proposed system
maintains near-$99\%$ classification accuracy over link lengths up to
$800$~km across $16$-, $64$-, and $256$-QAM, while the conventional JPEG
baselines exhibit pronounced reach limitations and abrupt performance
collapse as channel conditions worsen. At the same time, the semantic
system transmits each image using only $128$ symbols in a fixed
$4.0$~ns, reducing the transmission cost with respect to the
LDPC-coded JPEG baseline at every tested modulation order. 
The
results show that semantic communication can extend the usable reach of
nonlinear optical links while reducing the transmission resources
required to accomplish the target task. 

\bibliographystyle{IEEEtran}
\bibliography{references}

@article{Lan2021WhatIs,
author = {Lan, Qiao and others},
title = {What is Semantic Communication? A View on Conveying Meaning in the Era of Machine Intelligence},
journal = {Journal of Communications and Information Networks},
volume = {6},
number = {4},
pages = {336--371},
year = {2021},
doi = {10.23919/jcin.2021.9663101}
}

@article{Xie2021DeepLearning,
author = {Xie, Huiqiang and others},
title = {Deep Learning Enabled Semantic Communication Systems},
journal = {IEEE Transactions on Signal Processing},
volume = {69},
pages = {2663--2675},
year = {2021},
doi = {10.1109/tsp.2021.3071210}
}

@article{Qin2021Semantic,
author = {Qin, Zhijin and others},
title = {Semantic Communications: Principles and Challenges},
journal = {arXiv preprint arXiv:2201.01389},
year = {2021},
doi = {10.48550/arxiv.2201.01389}
}

@article{Zhang2022Deep,
author = {Zhang, Hongwei and others},
title = {Deep Learning-Enabled Semantic Communication Systems With Task-Unaware Transmitter and Dynamic Data},
journal = {IEEE Journal on Selected Areas in Communications},
volume = {40},
number = {12},
pages = {3229--3243},
year = {2022},
doi = {10.1109/jsac.2022.3221991}
}

@article{Gunduz2022Beyond,
author = {G{"u}nd{"u}z, Deniz and others},
title = {Beyond Transmitting Bits: Context, Semantics, and Task-Oriented Communications},
journal = {arXiv preprint arXiv:2207.09353},
year = {2022},
doi = {10.48550/arxiv.2207.09353}
}

@article{Shannon1948Mathematical,
author = {Shannon, C. E.},
title = {A Mathematical Theory of Communication},
journal = {Bell System Technical Journal},
volume = {27},
number = {3},
pages = {379--423},
year = {1948},
doi = {10.1002/j.1538-7305.1948.tb01338.x}
}

@article{Bourtsoulatze2019Deep,
author = {Bourtsoulatze, Eirina and others},
title = {Deep Joint Source-Channel Coding for Wireless Image Transmission},
journal = {IEEE Transactions on Cognitive Communications and Networking},
volume = {5},
number = {3},
pages = {567--579},
year = {2019},
doi = {10.1109/tccn.2019.2919300}
}

@inproceedings{Kurka2020Joint,
author = {Burth Kurka, David and G{"u}nd{"u}z, Deniz},
title = {Joint Source-Channel Coding of Images with (not very) Deep Learning},
booktitle = {2020 International Zurich Seminar on Information and Communication (IZS 2020)},
pages = {90--94},
year = {2020},
doi = {10.3929/ethz-b-000402967}
}

@book{Agrawal2019Nonlinear,
author = {Agrawal, Govind P.},
title = {Nonlinear Fiber Optics},
publisher = {Academic Press},
edition = {6th},
year = {2019},
doi = {10.1016/C2017-0-01119-4}
}

@article{Jovanovic2023Geometric,
author = {Jovanovic, Ognjen and others},
title = {Geometric Constellation Shaping for Fiber-Optic Channels via End-to-End Learning},
journal = {Journal of Lightwave Technology},
volume = {41},
number = {12},
pages = {3726--3736},
year = {2023},
doi = {10.1109/jlt.2023.3276300}
}

@article{Jovanovic2022End,
author = {Jovanovic, Ognjen and others},
title = {End-to-End Learning of a Constellation Shape Robust to Channel Condition Uncertainties},
journal = {Journal of Lightwave Technology},
volume = {40},
number = {10},
pages = {3316--3324},
year = {2022},
doi = {10.1109/jlt.2022.3169993}
}

@article{Gaiarin2020End,
author = {Gaiarin, Simone and others},
title = {End-to-End Optimization of Coherent Optical Communications Over the Split-Step Fourier Method Guided by the Nonlinear Fourier Transform Theory},
journal = {Journal of Lightwave Technology},
volume = {39},
number = {2},
pages = {418--428},
year = {2020},
doi = {10.1109/jlt.2020.3033624}
}

@article{Cai2026Convolutional,
author = {Cai, Xiaomin and others},
title = {Convolutional Autoencoder-Enhanced Semantic Communication in Optical Fiber Systems},
journal = {IEEE Transactions on Cognitive Communications and Networking},
year = {2026},
doi = {10.1109/tccn.2026.3677171}
}

@inproceedings{Cai2025Machine,
author = {Cai, Xiaomin and others},
title = {Machine Learning-Enhanced Semantic Communication in Optical Fiber Systems},
booktitle = {2025 25th Anniversary International Conference on Transparent Optical Networks (ICTON)},
pages = {1--4},
year = {2025},
doi = {10.1109/icton67126.2025.11125027}
}

@article{Poggiolini2026Polynomial,
author = {Poggiolini, P. and others},
title = {Polynomial Closed Form Model for Ultra-Wideband Transmission Systems},
journal = {Journal of Lightwave Technology},
year = {2026},
doi = {10.1109/jlt.2026.3678322}
}

@inproceedings{AitAoudia2018End,
author = {Ait Aoudia, Faycal and Hoydis, Jakob},
title = {End-to-End Learning of Communications Systems Without a Channel Model},
booktitle = {2018 52nd Asilomar Conference on Signals, Systems, and Computers},
pages = {298--303},
year = {2018},
doi = {10.1109/acssc.2018.8645416}
}

@article{Marcuse1997Application,
author = {Marcuse, D. and others},
title = {Application of the Manakov-{PMD} equation to studies of signal propagation in optical fibers with randomly varying birefringence},
journal = {Journal of Lightwave Technology},
volume = {15},
number = {9},
pages = {1735--1746},
year = {1997},
doi = {10.1109/50.622902}
}

@misc{opticommpy,
author       = {Forestieri, Enrico and others},
title        = {{OptiCommPy}: Open-source {P}ython library for optical communications},
year         = {2023},
howpublished = {\url{https\://github.com/edsonportosilva/OptiCommPy}},
}

@article{Savory2010Digital,
author = {Savory, Seb J.},
title = {Digital Coherent Optical Receivers: Algorithms and Subsystems},
journal = {IEEE Journal of Selected Topics in Quantum Electronics},
volume = {16},
number = {5},
pages = {1164--1179},
year = {2010},
doi = {10.1109/JSTQE.2010.2044751}
}

@inproceedings{Jang2017Categorical,
author = {Jang, Eric and others},
title = {Categorical Reparameterization with {G}umbel-{S}oftmax},
booktitle = {International Conference on Learning Representations},
year = {2017}
}

@article{Williams1992Simple,
author = {Williams, Ronald J.},
title = {Simple Statistical Gradient-Following Algorithms for Connectionist Reinforcement Learning},
journal = {Machine Learning},
volume = {8},
pages = {229--256},
year = {1992},
doi = {10.1007/BF00992696}
}

@article{Wang2004ImageQuality,
author = {Wang, Zhou and others},
title = {Image Quality Assessment: From Error Visibility to Structural Similarity},
journal = {IEEE Transactions on Image Processing},
volume = {13},
number = {4},
pages = {600--612},
year = {2004},
doi = {10.1109/TIP.2003.819861}
}

@article{lecun1998gradient,
author = {LeCun, Yann and others},
title = {Gradient-Based Learning Applied to Document Recognition},
journal = {Proceedings of the IEEE},
volume = {86},
number = {11},
pages = {2278--2324},
year = {1998},
doi = {10.1109/5.726791}
}

@article{Richardson2001Efficient,
author = {Richardson, Thomas J. and Urbanke, R{"u}diger L.},
title = {Efficient Encoding of Low-Density Parity-Check Codes},
journal = {IEEE Transactions on Information Theory},
volume = {47},
number = {2},
pages = {638--656},
month = feb,
year = {2001}
}

@inproceedings{He2016Deep,
  author    = {He, Kaiming and Zhang, Xiangyu and Ren, Shaoqing and Sun, Jian},
  title     = {Deep Residual Learning for Image Recognition},
  booktitle = {Proc. IEEE Conf. Comput. Vis. Pattern Recognit. (CVPR)},
  year      = {2016},
  pages     = {770--778},
  doi       = {10.1109/CVPR.2016.90}
}

\end{document}